\documentclass[12pt]{article}%
\usepackage{amssymb}
\usepackage{amsmath}
\usepackage{amsfonts}
\usepackage{graphicx}%
\providecommand{\U}[1]{\protect\rule{.1in}{.1in}}

\begin{document}
\bigskip%
\begin{titlepage}
\vspace{.3cm} \vspace{1cm}
\begin{center}
\baselineskip=16pt \centerline{\bf{\Large
{Mimetic Massive Gravity: Beyond Linear Approximation}}}
\vspace{1truecm}
\centerline{\large\bf Ali H.
Chamseddine$^{1,2}$\ , Viatcheslav Mukhanov$^{3,4,5}$\ \ } \vspace{.5truecm}
\emph{\centerline{$^{1}%
$Physics Department, American University of Beirut, Lebanon}}
\emph{\centerline{$^{2}$I.H.E.S. F-91440 Bures-sur-Yvette, France}}
\emph{\centerline{$^{3}%
$Theoretical Physics, Ludwig Maxmillians University,Theresienstr. 37, 80333 Munich, Germany }%
}
\emph{\centerline{$^{4}%
$MPI for Physics, Foehringer Ring, 6, 80850, Munich, Germany}}
\emph{\centerline{$^{5}%
$LPT de l'Ecole Normale Superieure, 24 rue Lhomond, 75231 Paris cedex, France}%
}
\end{center}
\vspace{2cm}
\begin{center}
{\bf Abstract}
\end{center}
We present a theory of ghost-free massive gravity where the mass of the graviton is generated through the
Brout-Englert-Higgs (BEH) mechanism and  one of the four scalar fields used is that of mimetic gravity. The mass term
is not of the Fierz-Pauli type and the constraint eliminates the Boulware-Deser ghost which is absent to all orders.
We perform a detailed analysis using the methods of cosmological perturbation theory and consider quantum
fluctuations of the degrees of freedom of massive graviton and mimetic matter. It is shown that for three of the degrees
of freedom of the graviton of mass $m$ the nonlinear corrections become comparable to the linear terms already at a
length scale of order $m^{-\frac{1}{2}}%
$. Thus, at smaller scales they become strongly coupled and the graviton
remains with two transverse degrees of freedom which get strongly coupled only at Planck scale. The mimetic field
behaves as cold particles of half of the graviton mass and could well explain the source of dark matter in our
universe. In the weakly coupled domain mimetic matter is completely decoupled from the massive graviton.
\end{titlepage}%

\section{Introduction}

The problem of finding a consistent theory of massive gravity has recently
attracted considerable interest. It is by now well established that the
simplest way of giving mass to the graviton without explicit breaking of
diffeomorphism invariance is by employing four scalar fields, which acquire
vacuum expectation values \cite{tH,CM}. As a result the scalar fields are
absorbed making the graviton massive on Minkowski background. However, four
scalar fields have generically four degrees of freedom and only three of them
are needed to provide mass to the graviton. Out of four scalar fields, used to
preserve global Lorentz invariance, one field must be taken with negative
kinetic energy and if its perturbations around a spontaneously broken symmetry
background propagates then there would appear a ghost mode leading to
inconsistency of the theory. By adopting the Fierz-Pauli (FP) term for the
mass \cite{FP} one can vanish the kinetic term for these perturbations, so
that the dangerous mode disappears in the linear approximation. Generically it
reappears again and starts to propagate on a nontrivial background and is
known as nonlinear Boulware-Deser (BD) ghost \cite{BD}. There was recently an
attempt to extend the FP-term to higher orders in such a way as to avoid the
propagation of this field to all orders on a nontrivial background. The
resulting theory which might avoid the non-linear BD-ghost is rather
unambiguous and its action is given by an infinite expansion of a square root
function \cite{Kab} which can be simplified and rewritten as a quadratic one
by making use of auxiliary fields \cite{CMquad}. However, in some particular
backgrounds, the ghost mode nevertheless gets excited, raising a question
about consistency of the theory \cite{CMghost,CMghost1}.

One can wonder whether it is possible to have a ghost-free massive gravity
which is not very restricted and well behaved to all higher orders? We will
show in this paper that this can be easily achieved in mimetic gravity
\cite{CMmim}, with non Fierz-Pauli mass term, where we use the constrained
scalar field imitating Dark Matter as one of the four fields needed to realize
Brout-Englert-Higgs (BEH) mechanism for the graviton mass.

\section{BEH-mechanism for massive gravity}

In BEH-mechanism for massive gravity a central role is played by four
\textit{scalar} fields $\phi^{A}\left(  x^{\mu}\right)  ,$ $A=0,1,2,3,$ which
in the broken symmetry phase, in Minkowski space-time, acquire vacuum
expectation values
\begin{equation}
\left\langle \phi^{A}\right\rangle =\delta_{\mu}^{A}x^{\mu}\equiv x^{A}.
\label{1}%
\end{equation}
Note that the vacuum state is degenerate and the different vacua are related
by Poicnare transformations. From the auxiliary induced metric $H^{AB}%
=g^{\mu\nu}\phi_{,\mu}^{A}\phi_{,\nu}^{B},$ where $\phi_{,\mu}^{A}\equiv
\frac{\partial\phi^{A}}{\partial x^{\mu}},$ it is convenient to build the
diffeomorphism invariant set of scalars%
\begin{equation}
\bar{h}^{AB}=H^{AB}-\eta^{AB}, \label{2}%
\end{equation}
and use them to give mass to the graviton. Here $\eta^{AB}=(1,-1,-1,-1)$ is
the auxiliary Minkowski metric. Let us consider the small perturbations of the
scalar fields around broken symmetry phase%
\begin{equation}
\phi^{A}=x^{A}+\chi^{A}, \label{3}%
\end{equation}
which induce small metric perturbations%
\begin{equation}
g^{\mu\nu}=\eta^{\mu\nu}+h^{\mu\nu}. \label{4}%
\end{equation}
Then
\begin{equation}
\bar{h}^{AB}=h^{AB}+\partial^{A}\chi^{B}+\partial^{B}\chi^{A}+h^{A\nu}%
\chi_{,\nu}^{B}+h^{B\nu}\chi_{,\nu}^{A}+\eta^{\mu\nu}\chi_{,\mu}^{A}\chi
_{,\nu}^{B}+h^{\mu\nu}\chi_{,\mu}^{A}\chi_{,\nu}^{B}, \label{5}%
\end{equation}
where $h^{AB}\equiv\delta_{\mu}^{A}\delta_{\nu}^{B}h^{\mu\nu},$ $\partial
^{A}\chi^{B}\equiv\eta^{A\nu}\chi_{,\nu}^{B},$ $\eta^{A\nu}\equiv\delta_{\mu
}^{A}\eta^{\mu\nu},$ $h^{A\nu}=\delta_{\mu}^{A}h^{\mu\nu}$ etc. The scalars
$\bar{h}^{AB}$ are diffeomorphism invariant and by an appropriate choice of
coordinate system $x^{\mu}$ we can impose four gauge conditions on fourteen
functions $\chi^{A}$ and $h^{\mu\nu}$ out of which they are built. For
instance, in so called \emph{unitary}\ gauge where $\chi^{A}=0,$ $\bar{h}%
^{AB}$ coincides with the metric perturbations $h^{\mu\nu}.$ Thus, it is clear
that $\overline{h}^{AB}$ can be used to construct, in a gauge invariant way,
the mass term for the graviton via BEH-mechanism. For instance, a theory with
the action%
\begin{equation}
S=\int d^{4}x\sqrt{-g}\left[  -\frac{1}{2}R+\frac{m^{2}}{8}\left(  \bar{h}%
^{2}-\bar{h}^{AB}\bar{h}_{AB}\right)  \right]  , \label{6}%
\end{equation}
where $\bar{h}=\bar{h}_{A}^{A}$ and we use the units in which $8\pi G=1$,
describes massive gravity with FP-mass term in broken symmetry phase. Note
that contraction of capital indices guarantees the invariance of the theory
with respect to a particular choice of the vacuum in a huge landscape of
degenerate vacuum states related by Poincare transformations.

One can easily see why the Fierz-Pauli combination is so special by
considering a decoupling limit when the metric perturbations $h^{\mu\nu}$
vanish. In this case the mass term
\begin{equation}
\frac{m^{2}}{8}\left(  \alpha\cdot\bar{h}^{2}-\bar{h}^{AB}\bar{h}_{AB}\right)
, \label{7}%
\end{equation}
where $\alpha$ is a numerical coefficient, becomes (up to total derivative)%
\begin{equation}
\frac{m^{2}}{2}\left(  -\frac{1}{4}F^{AB}F_{AB}+\left(  \alpha-1\right)
\left(  \partial_{A}\chi^{A}\right)  ^{2}+O\left(  \chi^{3}\right)  \right)
\label{8}%
\end{equation}
with $F_{AB}=\partial_{A}\chi_{B}-\partial_{B}\chi_{A}$. It is clear that
unless $\alpha=1,$ the theory with Lagrangian (\ref{8}) describes four fields
one of which is inevitably a ghost. Only the choice of the mass combination
$\bar{h}^{2}-\bar{h}^{AB}\bar{h}_{AB}$ insures that the term $\left(  \bar
{h}_{00}\right)  ^{2},$ that provides dynamics to the ghost field $\chi^{0},$
is absent in the Lagrangian and $\chi^{0}$ becomes similar to the gauge
potential $A^{0}$ in Maxwell theory, which corresponds to the first term in
(\ref{8}). However, if one considers the propagation of the fields in a curved
background (for instance induced by an external source) with metric $g_{\mu
\nu}=\eta_{\mu\nu}+h_{\mu\nu}$, then the contribution
\begin{equation}
\bar{h}^{2}=2\bar{h}_{i}^{i}\bar{h}_{0}^{0}+...=2h_{i}^{i}\left(  \dot{\chi
}^{0}\right)  ^{2}+..., \label{8a}%
\end{equation}
due to the higher order terms in (\ref{5}) induces a propagator for $\chi^{0}%
$, thus resurrecting the nonlinear BD-ghost in non-trivial background even for
FP mass term. One can try to get rid of the nonlinear ghost by adding to the
action (\ref{6}) higher order terms $\bar{h}^{3},\bar{h}\bar{h}^{AB}\bar
{h}_{AB},...$ and such strategy leads to a nearly unambiguous theory with
action \cite{Kab}:%
\begin{equation}
S=\int d^{4}x\sqrt{g}\left[  -\frac{1}{2}R+\frac{m^{2}}{2}\left(  S^{2}%
-S^{AB}S_{AB}\right)  \right]  , \label{8b}%
\end{equation}
where%
\begin{equation}
S_{AB}=\sqrt{\eta_{AB}+\bar{h}_{AB}}-\eta_{AB}. \label{8c}%
\end{equation}
To second order in $\overline{h}$ this theory reduces to FP theory. In higher
orders it is represented by an infinite series in $\bar{h}$ and only with the
help of auxiliary \emph{vierbein type}\ fields the square root in (\ref{8c})
can be given in finite form and the theory becomes quadratic in $S$
\cite{CMquad}. The theory (\ref{8b}) looks promising from the point of view of
keeping the field $\chi^{0}$ non-dynamical to all orders. However, it was
shown in \cite{CMghost}, \cite{CMquad} that the term $\bar{h}^{0i}\bar{h}%
_{0i},$ which is already present in FP-term, as part of the quadratic term
$h^{A\nu}\chi_{,\nu}^{B}$ in (\ref{5}), \ gives the following contribution to
the action%
\begin{equation}
\bar{h}^{0i}\bar{h}_{0i}=h^{0i}h^{0i}\left(  \dot{\chi}^{0}\right)  ^{2}+...
\label{9}%
\end{equation}
which is not canceled by higher order terms. Thus, in some backgrounds the
field $\chi^{0}$ starts to propagate again and there is no guarantee that this
mode would not induce a ghost in an arbitrary background. A reader could
wonder whether it is possible to choose a gauge $h^{0i}=0$ where the term
(\ref{9}) is absent$.$ It was shown in \cite{CMghost}, \cite{CMghost1} that in
this gauge the propagating $\chi^{0}$ mode reappears in a more complicated
way. This is why the question about ghost free massive gravity remains yet
open. In this paper we address the following questions:

\begin{itemize}
\item Is it in principle possible to construct ghost free theory which will
describe massive graviton with five degrees of freedom in Minkowski background ?

\item Must the FP mass term be necessarily always used in massive gravity?
\end{itemize}

\section{Mimetic \textit{non} Fierz-Pauli massive gravity}

As we have seen above the problem in massive gravity is that the field
$\phi^{0}$ becomes dynamical by itself and thus inevitably leads to the
appearance of linear or nonlinear ghost. The idea we exploit in this paper is
to use in BEH-mechanism the mimetic field \cite{CMmim} as one of four fields
responsible for the appearance of the graviton mass. This field $\phi^{0}$
always obeys the constraint
\begin{equation}
g^{\mu\nu}\partial_{\mu}\phi^{0}\partial_{\nu}\phi^{0}=1,\label{10}%
\end{equation}
and in a synchronous coordinate system serves as time coordinate, so that the
generic solution of (\ref{10}) is%
\begin{equation}
\phi^{0}=t+A\label{11}%
\end{equation}
where $A$ is a constant of integration. Combined with the longitudinal mode of
gravity it leads to the appearance of mimetic matter which can well imitate
the observed Dark Matter in the Universe \cite{CMmim}, \cite{CMVmim}. The
remaining three scalar fields can provide three extra degree of freedom needed
for the massive graviton. Thus, in such theory we are guaranteed to have no
ghosts to any order in perturbation theory. It happens that in this case the
mass term for the graviton is unambiguously fixed. It is not of the
Fierz-Pauli type and the action of the corresponding theory is
\begin{equation}
S=%
{\displaystyle\int}
d^{4}x\sqrt{g}\left[  -\frac{1}{2}R+\frac{m^{2}}{8}\left(  \frac{1}{2}\bar
{h}^{2}-\bar{h}^{AB}\bar{h}_{AB}\right)  +\lambda\left(  g^{\mu\nu}%
\partial_{\mu}\phi^{0}\partial_{\nu}\phi^{0}-1\right)  \right]  .\label{12}%
\end{equation}
The mass term has a relative coefficient of $-\frac{1}{2}$ between the
$\bar{h}^{2}$ and $\bar{h}^{AB}\bar{h}_{AB}$ terms. The mimetic constraint
term does not violate the Lorentz invariance of the ground state because it is
obviously invariant with respect to the transformation $\phi^{0}%
\rightarrow\tilde{\phi}^{0}=\Lambda_{B}^{0}\phi^{B}$ for $\phi^{B}=\delta
_{\mu}^{B}x^{\mu}.$ Variation of this action with respect to the metric
$g_{\mu\nu}$ gives the following modified Einstein equations
\begin{align}
G_{\nu}^{\mu} &  =2\lambda\partial^{\mu}\phi^{0}\partial_{\nu}\phi^{0}%
-\frac{m^{2}}{8}\left(  \frac{1}{2}\bar{h}^{2}-\bar{h}^{AB}\bar{h}%
_{AB}\right)  \delta_{\nu}^{\mu}\nonumber\\
&  +\frac{m^{2}}{2}\left(  \frac{1}{2}\bar{h}\partial^{\mu}\phi_{A}%
\partial_{\nu}\phi^{A}-\bar{h}_{B}^{A}\partial^{\mu}\phi_{A}\partial_{\nu}%
\phi^{B}\right)  ,\label{13}%
\end{align}
where $G_{\nu}^{\mu}=R_{\nu}^{\mu}-\frac{1}{2}R\delta_{\nu}^{\mu}$ is the
Einstein tensor and we recall that capital indices are raised and lowered with
the help of the auxiliary metric $\eta_{AB},$ while for Greek indices the
space-time metric $g_{\mu\nu}$ is used. Variation with respect to the scalar
fields $\phi_{A}$ leads to%
\begin{equation}
\partial_{\mu}\left(  \sqrt{g}\left[  \frac{m^{2}}{2}\left(  \frac{1}{2}%
\bar{h}\partial^{\mu}\phi^{A}-\bar{h}_{B}^{A}\partial^{\mu}\phi^{B}\right)
+2\lambda\delta_{0}^{A}\partial^{\mu}\phi^{0}\right]  \right)  =0.\label{phi}%
\end{equation}
Finally the constraint (\ref{10}), which follows by varying with respect to
$\lambda,$ takes the simple form
\begin{equation}
\bar{h}_{0}^{0}=0.\label{14}%
\end{equation}
In the next section we will analyze linear perturbations in this model in
component form, using the methods of cosmological perturbation theory
\cite{vm}. An explicitly covariant study of these perturbations is given in a
separate publication \cite{PRL}.

\section{Graviton on Minkowski background}

We first study the linearized theory considering small metric perturbations in
Minkowski background, that is, assuming
\begin{equation}
g_{\mu\nu}=\eta_{\mu\nu}+h_{\mu\nu}, \label{15}%
\end{equation}
and $\left\vert h_{\mu\nu}\right\vert \ll1.$ Correspondingly the scalar fields
deviate from their vacuum expectations values by $\chi^{A},$ that is,%
\begin{equation}
\phi^{A}=x^{A}+\chi^{A}. \label{16}%
\end{equation}
To explicitly reveal the true physical degrees of freedom of the massive
graviton it is convenient to use the methods of cosmological perturbation
theory and classify the perturbations with respect to irreducible
representations of the spatial rotation group \cite{vm}. The metric component
$h_{00}$ behaves as a scalar under rotations and it is convenient to denote it
as%
\begin{equation}
h_{00}=2\phi, \label{17}%
\end{equation}
where $\phi$ is a 3-scalar. The space-time components $h_{0i}$ behaves as a
3-vector that can be written as a sum of a longitudinal \ and transverse
parts:%
\begin{equation}
h_{0i}=B_{,i}+S_{i}, \label{18}%
\end{equation}
where $B_{,i}=\frac{\partial B}{\partial x^{i}}$ and $S_{i}$ has zero
divergence, that is, $\partial^{i}S_{i}=0.$ Finally $h_{ij}$ can be decomposed
as%
\begin{equation}
h_{ij}=2\psi\delta_{ij}+2E_{,ij}+F_{i,j}+F_{j,i}+\tilde{h}_{ij}, \label{19}%
\end{equation}
with $\partial^{i}F_{i}=0$ and the transverse traceless part $\tilde{h}_{ij}$
satisfies four conditions $\partial^{i}\tilde{h}_{ij}=0,$ $\tilde{h}_{i}%
^{i}=0,$ leaving us with two polarizations for the massless graviton in
General Relativity$.$ Thus, the perturbations can be classified as
\textit{scalar} perturbations described by $\phi,$ $\psi,$ $B,$ and $E,$
\textit{vector} perturbations corresponding to $S_{i}$ and $F_{i}$ and
\textit{tensor} perturbations $\tilde{h}_{ij}.$ In the linear approximation
they are completely decoupled and thus can be studied separately$.$ In General
Relativity in empty space the scalar and vector perturbations vanish and they
are induced entirely by matter. In our theory, which can be treated as
Einstein theory with extra scalar fields, these perturbations are due to the
small perturbations $\chi^{A}$, which in turn can be decomposed as:%
\begin{equation}
\chi^{0}=\chi^{0},\text{ \qquad}\chi^{i}=\tilde{\chi}^{i}+\pi_{,i} \label{20}%
\end{equation}
where $\partial_{i}\tilde{\chi}^{i}=0.$ It is obvious that $\chi^{0}$ and
$\pi$ excite scalar modes, while $\tilde{\chi}^{i}$ is responsible for vector
perturbations. The remaining coordinate freedom%
\begin{equation}
x^{0}\rightarrow x^{0}+\xi^{0},\text{ \qquad\ }x^{i}\rightarrow x^{i}+\xi
^{i}=x^{i}+\zeta_{,i}+\tilde{\xi}^{i} \label{21}%
\end{equation}
with $\partial_{i}\tilde{\xi}^{i}=0,$ allows us to impose four gauge
conditions. Two of them, which are due to the freedom in choosing $\xi^{0}$
and $\zeta$ refer to the scalar perturbations sector, where for instance two
out of seven functions $\phi,\psi,B,$ $E,\chi^{0},\pi$ and $\lambda$ can be
taken to vanish. Often it is very convenient to impose the Newtonian gauge
conditions $B=E=0,$ which selects the coordinate system with explicit
Newtonian limit of General Relativity. The coordinate freedom due to
transverse $\tilde{\xi}^{i}$ leave the gauge freedom for the vector
perturbations and we can set either $S_{i}$ or $F_{i}$ to be equal to zero.
The transverse traceless part of the metric $\tilde{h}_{ij}$ is gauge invariant.

Keeping in equations (\ref{13}) only those terms which contain the first order
terms in perturbations, these equations are simplified to%
\begin{equation}
G_{0}^{0}\simeq2\lambda+\frac{m^{2}}{4}\bar{h}_{l}^{l}, \label{22a}%
\end{equation}%
\begin{equation}
G_{i}^{0}\simeq-\frac{m^{2}}{2}\bar{h}_{i}^{0}, \label{23}%
\end{equation}%
\begin{equation}
G_{k}^{i}\simeq\frac{m^{2}}{2}\left(  \frac{1}{2}\bar{h}_{l}^{l}\delta_{k}%
^{i}-\bar{h}_{k}^{i}\right)  , \label{24}%
\end{equation}
where we took into account that $\lambda,$ which is due to mimetic matter, is
a first order quantity in perturbations and $\bar{h}_{0}^{0}=0$ thanks to
constraint (\ref{14}).

\bigskip

\textbf{Scalar perturbations. }We fix the gauge by taking $B=E=0.$ In this
gauge the metric takes the form%
\begin{equation}
ds^{2}=(1+2\phi)dt^{2}-\left(  1-2\psi\right)  \delta_{ik}dx^{i}dx^{k},
\label{25}%
\end{equation}
where $\phi$ is the Newtonian gravitational potential. As follows from
(\ref{5}) we have to first order in perturbations
\begin{equation}
\bar{h}_{0}^{0}=-2\phi+2\dot{\chi}^{0},\qquad\text{ }\bar{h}_{i}^{0}=\chi
_{,i}^{0}-\dot{\pi}_{,i},\text{\qquad}\bar{h}_{k}^{i}=2\psi\delta_{k}^{i}%
+2\pi_{,ij}, \label{26}%
\end{equation}
where dot denotes derivative with respect to time $t.$ Taking into account
that in this gauge (see, \cite{vm}):
\begin{equation}
G_{0}^{0}=2\Delta\psi,\text{ }G_{i}^{0}=2\dot{\psi}_{,i},\text{ }G_{k}%
^{i}=\left(  \phi-\psi\right)  _{,ik}-(2\ddot{\psi}+\Delta\left(  \phi
-\psi\right)  )\delta_{ik}, \label{27}%
\end{equation}
and substituting the expressions above into (\ref{22a})-(\ref{24}) we obtain%
\begin{equation}
\Delta\psi=\lambda+\frac{m^{2}}{4}\left(  3\psi+\Delta\pi\right)  , \label{28}%
\end{equation}%
\begin{equation}
\dot{\psi}=-\frac{m^{2}}{4}\left(  \chi^{0}-\dot{\pi}\right)  , \label{29}%
\end{equation}
from $0-0$ and $0-i$ equations. The off diagonal $i\neq k$ \ components of
(\ref{24}) give%
\begin{equation}
\phi-\psi=-m^{2}\pi, \label{30}%
\end{equation}
while taking the trace we obtain
\begin{equation}
3\ddot{\psi}+\Delta\left(  \phi-\psi\right)  +\frac{m^{2}}{4}\left(
3\psi+\Delta\pi\right)  =0. \label{31}%
\end{equation}
The constraint equation (\ref{14}), when linearized, takes the form%
\begin{equation}
\phi=\dot{\chi}^{0}, \label{32}%
\end{equation}
and together with equations (\ref{28})-(\ref{31}) are enough to determine the
five unknown variables $\phi,$ $\psi,$ $\lambda,$ $\chi^{0}$ and $\pi.$
Differentiating (\ref{29}) and using (\ref{32}) and (\ref{30}) we have%
\begin{equation}
\ddot{\psi}=-\frac{m^{2}}{4}\left(  \dot{\chi}^{0}-\ddot{\pi}\right)
=-\frac{m^{2}}{4}\left(  \psi-m^{2}\pi-\ddot{\pi}\right)  . \label{33}%
\end{equation}
Substituting this expression in (\ref{31}) and taking into account (\ref{30})
we derive the closed form equation for $\pi:$%
\begin{equation}
\ddot{\pi}-\Delta\pi+m^{2}\pi=0. \label{34}%
\end{equation}
Considering a plane-wave with wave-number $\vec{k}$ the solution of this
equation is
\begin{equation}
\pi_{k}=\left(  A_{k}\sin\omega t+B_{k}\cos\omega t\right)
e^{i\overrightarrow{k}\cdot\overrightarrow{x}},\qquad\omega^{2}=k^{2}+m^{2},
\label{35}%
\end{equation}
where $A_{k}$ and $B_{k}$ are constants of integration. Using (\ref{34}) we
can rewrite (\ref{33}) as%
\begin{equation}
\ddot{\psi}+\frac{m^{2}}{4}\psi=\frac{m^{2}}{4}\Delta\pi\label{36}%
\end{equation}
and substituting (\ref{35}) in the right hand side of this equation we solve
it to obtain%
\begin{align}
\psi &  =C\left(  x^{i}\right)  \sin\frac{mt}{2}+D\left(  x^{i}\right)
\cos\frac{mt}{2}\nonumber\\
&  +\frac{m^{2}k^{2}}{4k^{2}+3m^{2}}\left(  A_{k}\sin\omega t+B_{k}\cos\omega
t\right)  e^{i\overrightarrow{k}\cdot\overrightarrow{x}}, \label{37}%
\end{align}
where $C\left(  x^{i}\right)  $ and $D\left(  x^{i}\right)  $ are further
space dependent constants of integration describing the contribution of
mimetic matter to the gravitational potential $\psi.$ It then follows from
(\ref{30}) that%
\begin{align}
\phi &  =\psi-m^{2}\pi=C\left(  x^{i}\right)  \sin\frac{mt}{2}+D\left(
x^{i}\right)  \cos\frac{mt}{2}\nonumber\\
&  -\frac{3m^{2}\omega^{2}}{4k^{2}+3m^{2}}\left(  A_{k}\sin\omega t+B_{k}%
\cos\omega t\right)  e^{i\overrightarrow{k}\cdot\overrightarrow{x}}.
\label{38}%
\end{align}
From equation (\ref{29}) one obtains%
\begin{align}
\chi^{0}  &  =\dot{\pi}-\frac{4}{m^{2}}\dot{\psi}=-\frac{2}{m}\left(  C\left(
x^{i}\right)  \cos\frac{mt}{2}-D\left(  x^{i}\right)  \sin\frac{mt}{2}\right)
\nonumber\\
&  +\frac{3m^{2}\omega}{4k^{2}+3m^{2}}\left(  A_{k}\cos\omega t-B_{k}%
\sin\omega t\right)  e^{i\overrightarrow{k}\cdot\overrightarrow{x}}.
\label{39}%
\end{align}
Finally we solve for $\lambda$ from equation (\ref{28}) to get
\begin{equation}
\lambda=\left(  \triangle-\frac{3m^{2}}{4}\right)  \left(  C\left(
x^{i}\right)  \sin\frac{mt}{2}+D\left(  x^{i}\right)  \cos\frac{mt}{2}\right)
\label{40}%
\end{equation}
Having determined that the scalar mode of massive graviton is represented by
the field $\pi$ satisfying (\ref{34}), we note that $\lambda$ is entirely
decoupled from $\pi$ and obeys the equation
\begin{equation}
\ddot{\lambda}+\frac{m^{2}}{4}\lambda=0, \label{41}%
\end{equation}
which describes massive mimetic matter. The contribution of this mimetic
matter to the gravitational potentials is%
\begin{equation}
\phi=\psi=C\left(  x^{i}\right)  \sin\frac{mt}{2}+D\left(  x^{i}\right)
\cos\frac{mt}{2}. \label{42}%
\end{equation}
To find the normalized independent quantization variables we need to calculate
the action for the scalar perturbations. Let us first expand the action
(\ref{12}) to second order in scalar perturbations in the conformal Newtonian
gauge, where $B=E=0,$
\begin{align}
S  &  =%
{\displaystyle\int}
d^{4}x\left[  -3\dot{\psi}^{2}-\psi\triangle\psi+2\phi\triangle\psi
+2\lambda\left(  \dot{\chi}^{0}-\phi\right)  \right. \nonumber\\
&  -\frac{m^{2}}{4}\left(  \dot{\chi}^{0}-\phi\right)  ^{2}+\frac{m^{2}}%
{2}\left(  \dot{\chi}^{0}-\phi\right)  \left(  3\psi+2\triangle\pi\right)
\nonumber\\
&  \left.  +\frac{m^{2}}{4}\left(  3\psi^{2}+2\psi\triangle\pi-\triangle
\pi\triangle\pi\right)  +\frac{m^{2}}{4}\left(  \chi^{0}-\dot{\pi}\right)
_{,i}\left(  \chi^{0}-\dot{\pi}\right)  _{,i}\right]  \label{44}%
\end{align}
Variation of this action with respect to $\lambda$ gives the constraint
$\phi=\dot{\chi}^{0}$, which when substituted in the action (\ref{44}) reduces
it to%
\begin{align}
S  &  =%
{\displaystyle\int}
d^{4}x\left[  -3\dot{\psi}^{2}-\psi\triangle\psi+2\dot{\chi}^{0}\triangle
\psi\right. \nonumber\\
&  \left.  +\frac{m^{2}}{4}\left(  3\psi^{2}+2\psi\triangle\pi-\triangle
\pi\triangle\pi+\left(  \chi^{0}-\dot{\pi}\right)  _{,i}\left(  \chi^{0}%
-\dot{\pi}\right)  _{,i}\right)  \right]  \label{45}%
\end{align}
Next, as a result of variation with respect to $\chi^{0}$ we obtain the
constraint (\ref{29}), and this allows us to express $\chi^{0}$ and $\dot
{\chi}^{0}$ in terms of $\psi$ and $\pi,$%
\begin{equation}
S=%
{\displaystyle\int}
d^{4}x\left[  \dot{\psi}\left(  -3+\frac{4\triangle}{m^{2}}\right)  \dot{\psi
}-\psi\triangle\psi-2\dot{\pi}\triangle\dot{\psi}+\frac{m^{2}}{4}\left(
3\psi^{2}+2\psi\triangle\pi-\triangle\pi\triangle\pi\right)  \right]
\label{46}%
\end{equation}
The fields $\psi$ and $\pi$ have mixed propagators. To diagonalize this action
we substitute
\begin{equation}
\psi=\left(  \triangle-\frac{3m^{2}}{4}\right)  ^{-1}\left(  \lambda
+\frac{m^{2}}{4}\triangle\pi\right)  , \label{47}%
\end{equation}
that follows from $0-0$ equation (\ref{28}). After some algebra one finds that
the action separates in terms of $\pi$ and $\lambda$,
\begin{align}
S  &  =-\frac{1}{2}%
{\displaystyle\int}
d^{4}x\triangle\pi\left(  \frac{2m^{2}}{3m^{2}-4\triangle}\right)  \left(
\partial_{0}^{2}-\triangle+m^{2}\right)  \triangle\pi+\nonumber\\
&  +\frac{1}{2}%
{\displaystyle\int}
d^{4}x\lambda\left(  \frac{32}{m^{2}\left(  3m^{2}-4\triangle\right)
}\right)  \left(  \partial_{0}^{2}+\frac{m^{2}}{4}\right)  \lambda, \label{48}%
\end{align}
where one should understand $\triangle$ to be $-k^{2}$ for the plane-wave
modes with the wave-number $\vec{k}.$ This shows that the correctly normalized
modes of the corresponding quantum fields are
\begin{equation}
\sqrt{\frac{2m^{2}}{3m^{2}+4k^{2}}}k^{2}\pi_{k},\text{ \ \qquad\ }\sqrt
{\frac{32}{m^{2}\left(  3m^{2}+4k^{2}\right)  }}\lambda_{k}, \label{49}%
\end{equation}
for the graviton and mimetic matter. The typical amplitude of minimal quantum
fluctuations of the properly normalized fields (\ref{49}) in scales
$L\sim\frac{1}{k}$ is of order $\left(  \frac{k^{3}}{\omega_{k}}\right)
^{\frac{1}{2}}$ (see, for instance, formula (4.34) in \cite{MW}). Therefore,
taking into account that $\omega_{k}=\sqrt{k^{2}+m^{2}}$ for $\pi$ mode and
$\omega_{k}=\frac{1}{2}m$ for $\lambda$ mode we infer from (\ref{49}) that the
minimal level of quantum fluctuations in scales $L\ll m^{-1}$ for $\pi$ and
$\lambda$ are
\begin{equation}
\delta\pi_{L}\simeq m^{-1},\text{ \ \qquad}\delta\lambda_{L}\simeq\left(
\frac{m}{L^{5}}\right)  ^{\frac{1}{2}}. \label{50}%
\end{equation}
As it follows from (\ref{47}) and (\ref{30}) they induce the corresponding
metric perturbations of order%
\begin{equation}
\delta\phi_{L}\sim\delta\psi_{L}\simeq m^{2}\delta\pi_{L}\simeq m, \label{30a}%
\end{equation}
and%
\begin{equation}
\delta\phi_{L}=\delta\psi_{L}\simeq L^{2}\delta\lambda_{L}\simeq\left(
\frac{m}{L}\right)  ^{\frac{1}{2}}, \label{30b}%
\end{equation}
for the scalar graviton mode and mimetic matter, respectively. Thus the metric
perturbations due to the quantum fluctuations of the scalar modes always
remain much smaller than unity up to the Planckian scales. Note that we work
in Planck units where all constants are set to unity and in dimensional units
the right hand side, for instance in (\ref{30b}), is $\left(  \frac
{m}{m_{\mathrm{Pl}}}\right)  ^{\frac{1}{2}}\left(  \frac{l_{\mathrm{Pl}}}%
{L}\right)  ^{\frac{1}{2}}$, where $m_{\mathrm{Pl}}$ and $l_{\mathrm{Pl}}$ are
the Planck mass and Planck length respectively.

The sign of kinetic energy for the field $\pi$ is positive and this field is
not a ghost, while, as follows from (\ref{48}), the apparent contribution to
the \textit{energy density} of mimetic field from the mode with $k\gg m,$
\begin{equation}
-\frac{4}{m^{2}k^{2}}\left(  \dot{\lambda}^{2}+\frac{m^{2}}{4}\lambda
^{2}\right)  , \label{51}%
\end{equation}
is negative and looks singular as $m^{2}\rightarrow0.$ However, a propagator
for $\lambda$ does not not include a Laplacian. Therefore, $\dot{\lambda
}\propto m\lambda$, and hence the singularity $\frac{1}{m^{2}}$ in (\ref{51})
is canceled$.$ Moreover, for the mimetic matter the main contribution to the
energy density is linear in $\lambda,$ so that the total energy density is%
\begin{equation}
\varepsilon_{mim}\simeq\lambda-\frac{\lambda^{2}}{k^{2}}, \label{52}%
\end{equation}
where the second negative term just account for the negative contribution of
gravitational self-interaction to the total energy density. This second term
is smaller than the first term for $\lambda\ll k^{2}.$ When $\lambda$ becomes
of order $k^{2}$ then we immediately see from (\ref{49}) that $\psi$ becomes
of order unity and linear perturbation theory breaks down. To clarify the
situation further let us consider in Minkowski space\ a ball of radius $R$
filled by dust at rest with local energy density $\lambda.$ Then the
gravitational potential far away from the ball is determined by the mass of
the ball which in the leading order can be estimated as $M_{0}\simeq\lambda
R^{3}.$ However, in the next order, the negative gravitational energy of self
interaction of matter within the ball, which is of order $-\frac{M_{0}^{2}}%
{R},$ also contributes to the mass observed far away from the ball. The
resulting mass measured by a far away observer is%
\begin{equation}
M_{\mathrm{obs}}\simeq M_{0}-\frac{M_{0}^{2}}{R}. \label{53}%
\end{equation}
If the observer would \textrm{forget}\ about contribution of negative
gravitational energy to the total mass he would conclude that the energy
density within the ball is%
\begin{equation}
\varepsilon_{\mathrm{obs}}\simeq\frac{M_{\mathrm{obs}}}{R^{3}}\simeq
\lambda-R^{2}\lambda^{2}. \label{54}%
\end{equation}
Taking into account that $k\simeq R^{-1}$ this explains the origin of negative
energy contribution to the mimetic matter energy density. When the second term
in (\ref{54}) becomes comparable with the first one we first get semi-closed
worlds and finally when the the negative self-interaction energy exactly
compensates the internal mass one obtains the closed universe with zero total
energy with respect to an observer in asymptotically Minkowski space (see, for
example, \cite{ZN} for details).

\bigskip

\textbf{Vector perturbations. }Let us now turn to vector perturbations taking
the gauge $F_{i}=0$, so that the metric becomes%
\begin{equation}
ds^{2}=dt^{2}+2S_{i}dx^{i}dt-\delta_{ik}dx^{i}dx^{k}. \label{55}%
\end{equation}
As follows from (\ref{5}), (\ref{19}) and (\ref{20}), we have, to linear order
in perturbations,%
\begin{equation}
\bar{h}_{0}^{0}=0,\text{ \qquad}\bar{h}_{i}^{0}=-S_{i}+\dot{\tilde{\chi}}%
_{i},\text{ \qquad\ }\bar{h}_{k}^{i}=-\tilde{\chi}_{k,i}-\tilde{\chi}_{i,k},
\label{56}%
\end{equation}
where as we recall $\tilde{\chi}_{i}=-\tilde{\chi}^{i}$ and both $S_{i}$ and
$\tilde{\chi}_{i}$ are transverse. Taking into account that in the gauge
$F_{i}=0$:
\begin{equation}
G_{0}^{0}=0,\text{ \qquad}G_{i}^{0}=\frac{1}{2}\Delta S_{i},\text{ \qquad
}G_{k}^{i}=-\frac{1}{2}(\dot{S}_{i,k}+\dot{S}_{k,i}), \label{57}%
\end{equation}
(see, \cite{vm}) equations (\ref{23}) and (\ref{24}) reduce to
\begin{equation}
\Delta S_{i}=m^{2}\left(  S_{i}-\dot{\tilde{\chi}}_{i}\right)  , \label{58}%
\end{equation}%
\begin{equation}
\dot{S}_{i,k}+\dot{S}_{k,i}=-m^{2}\left(  \tilde{\chi}_{i,k}+\tilde{\chi
}_{k,i}\right)  , \label{59}%
\end{equation}
while equation (\ref{22a}) is satisfied identically. As follows from
(\ref{58})
\begin{equation}
S_{i}=\frac{m^{2}}{m^{2}-\Delta}\dot{\tilde{\chi}}_{i}, \label{60}%
\end{equation}
and this, after being substituted in (\ref{59}), gives
\begin{equation}
\left(  \ddot{\tilde{\chi}}_{i,k}+\ddot{\tilde{\chi}}_{k,i}\right)
-\Delta\left(  \tilde{\chi}_{i,k}+\tilde{\chi}_{k,i}\right)  +m^{2}\left(
\tilde{\chi}_{i,k}+\tilde{\chi}_{k,i}\right)  =0, \label{61}%
\end{equation}
or, equivalently,%
\begin{equation}
\ddot{\tilde{\chi}}_{i}-\Delta\tilde{\chi}_{i}+m^{2}\tilde{\chi}_{i}=0.
\label{62}%
\end{equation}
This equation describes two vector modes of the massive graviton. The action
for the vector modes can be obtained by expanding (\ref{12}) to second order
in perturbations:%
\begin{equation}
S=\frac{1}{4}\int d^{4}x\,\left[  S_{i,k}S_{i,k}+m^{2}\left(  \left(
\dot{\tilde{\chi}}_{i}-S_{i}\right)  \left(  \dot{\tilde{\chi}}_{i}%
-S_{i}\right)  -\tilde{\chi}_{i,k}\tilde{\chi}_{i,k}\right)  \right]  .
\label{63}%
\end{equation}
Variation of this action with respect to $S_{i}$ gives us constraint equation
(\ref{58}), and upon substituting (\ref{60}), the action reduces to%
\begin{equation}
S=-\frac{1}{2}\int d^{4}x\,\left[  \tilde{\chi}_{i}\left(  \frac{m^{2}\Delta
}{2\left(  \Delta-m^{2}\right)  }\right)  \left(  \partial_{0}^{2}%
-\Delta+m^{2}\right)  \tilde{\chi}_{i}\right]  . \label{64}%
\end{equation}
From this we deduce that the properly normalized modes for the vector
perturbations are
\begin{equation}
\sqrt{\frac{m^{2}\Delta}{2\left(  \Delta-m^{2}\right)  }}\tilde{\chi}_{i}.
\label{65}%
\end{equation}
Thus for plane-wave perturbations with momentum $k\simeq L^{-1}\gg m$ the
typical minimal quantum fluctuations in scales $L$ is of order%
\begin{equation}
\delta\tilde{\chi}_{L}\simeq\frac{1}{mL}. \label{66}%
\end{equation}
Taking into account that $\delta\dot{\tilde{\chi}}_{L}\simeq\frac{\delta
\tilde{\chi}_{L}}{L}$ we infer from (\ref{60}) that the corresponding metric
perturbations are%
\begin{equation}
\delta S_{iL}\simeq m \label{67}%
\end{equation}

\bigskip

\textbf{Tensor perturbations. }The tensor perturbations $\tilde{h}_{ij}$
satisfy four extra conditions $\partial^{i}\tilde{h}_{ij}=0=\tilde{h}_{i}%
^{i}.$ They are gauge invariant and describe two degree of freedom of the
graviton which have become massive. The equation for $\tilde{h}_{ij}$
immediately follows from (\ref{24}),
\begin{equation}
\left(  \partial_{0}^{2}-\Delta+m^{2}\right)  \tilde{h}_{ij}=0, \label{68}%
\end{equation}
and the corresponding action is%
\begin{equation}
S=-\frac{1}{8}\int d^{4}x\left[  \tilde{h}_{ij}\left(  \partial_{0}^{2}%
-\Delta+m^{2}\right)  \tilde{h}_{ij}\right]  . \label{69}%
\end{equation}
The canonical quantization variable is $\tilde{h}_{ij}$ and hence the typical
amplitude of quantum fluctuations for tensor modes in scales $L$ is of order%
\begin{equation}
\delta\tilde{h}_{ij}\simeq\frac{1}{L}, \label{70}%
\end{equation}
for $L\ll m^{-1}.$ They become of order unity at Planck scale $l_{\mathrm{Pl}%
}\simeq10^{-33}$ $\mathrm{cm}$ where they enter non-perturbative quantum
gravity strong coupling regime and linearized theory ceases to be applicable.
Although the amplitude of quantum scalar and vector metric perturbations is
scale independent $\delta\psi_{L}\simeq\delta S_{L}\simeq m$ and remains small
even at the Planck scale, these perturbations nevertheless enter the strong
coupling regime at the energy scale which is much below the Planck scale, that
is, well before the tensor modes (see next section).

Thus we have shown in this section that the theory with action (\ref{12})
describes on a Minkowski background a massive graviton with five degrees of
freedom. According to little (rotation) group representations they can be
thought of as consisting of one scalar degree of freedom described by $\pi
,\ $two vector degrees of freedom $\tilde{\chi}_{i}$ and two tensor modes
$\tilde{h}_{ij}.$ In addition we have a massive mimetic matter described by
$\lambda$ which, in the linear approximation, completely decouples from the
massive graviton. These results are valid only when the higher order
corrections to the linearized equation are negligible. In the next section we
determine the range of applicability of the linearized theory.

\section{Strong coupling and massless limit}

\textbf{Scalar modes.} To determine at which scales the nonlinear corrections
to the linearized equations become important we need to calculate the next
order terms in equations (\ref{13}). Because the resulting expressions are too
cumbersome even to second order, we will keep in the formulae only those terms
which can become comparable to linear terms assuming that the gravitational
potentials $\phi$ and $\psi$ are much smaller than unity and considering
perturbations with $k^{2}\gg m^{2}.$ In the left hand side of equations
(\ref{13}) it is enough to keep only the linear terms because under the
assumptions stated above all terms of order $\phi^{2},$ $\psi^{2}$ are small
compared to the linear terms. Keeping in $\bar{h}_{0}^{0},$ defined in
(\ref{5}), only terms to second order in perturbations we have
\begin{equation}
0=\bar{h}_{0}^{0}=-2\left(  \phi-\dot{\chi}^{0}\right)  +4\left(  \phi
-\frac{1}{2}\dot{\chi}^{0}\right)  ^{2}-\chi_{,i}^{0}\chi_{,i}^{0} \label{71}%
\end{equation}
It is clear that $O\left(  \phi^{2}\right)  $ terms are always much smaller
than $\phi$ and can be skipped. The term $4\phi\dot{\chi}^{0}$ is much smaller
than $\dot{\chi}^{0}$ for $\phi\ll1$ and hence can be neglected in $\bar
{h}_{0}^{0}.$ Therefore, keeping only relevant terms to second order in
perturbations, the constraint $\bar{h}_{0}^{0}=0$ becomes%
\begin{equation}
-2\phi+2\dot{\chi}^{0}+\left(  \dot{\chi}^{0}\right)  ^{2}-\chi_{,i}^{0}%
\chi_{,i}^{0}\simeq0. \label{72}%
\end{equation}
For linear perturbations we can express $\phi$, $\psi$ and $\chi^{0}$ in terms
of the independent fields $\pi$ and $\lambda$ using (\ref{35})-(\ref{40}):%
\begin{equation}
\phi=\dot{\chi}^{0}=-\frac{3m^{2}\omega^{2}\pi+4\lambda}{4k^{2}+3m^{2}}%
\simeq-\frac{3m^{2}}{4}\pi-\frac{\lambda}{k^{2}}, \label{73}%
\end{equation}%
\begin{equation}
\psi=\frac{m^{2}k^{2}\pi-4\lambda}{4k^{2}+3m^{2}}\simeq\frac{m^{2}}{4}%
\pi-\frac{\lambda}{k^{2}}, \label{74}%
\end{equation}%
\begin{equation}
\chi^{0}=\frac{3m^{4}\dot{\pi}+16\dot{\lambda}}{m^{2}\left(  4k^{2}%
+3m^{2}\right)  }\simeq\frac{3m^{2}}{4k}\pi+\frac{2}{mk^{2}}\lambda,
\label{75}%
\end{equation}
where to simplify the formulae in the second equality we considered
perturbations with scales $L\simeq k^{-1}\ll m^{-1}$ and therefore skipped all
subleading corrections proportional to $\frac{m^{2}}{k^{2}}.$ Moreover in
(\ref{75}) we estimated the time derivatives as $\dot{\lambda}\simeq\frac
{m}{2}\lambda$ and $\dot{\pi}\simeq k\pi.$ It is clear that the $\left(
\dot{\chi}^{0}\right)  ^{2}$ term in (\ref{72}) is of order $\phi^{2}$ and can
be neglected compared to the last term, which in the domain of applicability
of the linear theory would signal us when the linearized equations fail.
Namely, this happens at scales where $\chi_{,i}^{0}\chi_{,i}^{0}$ $\sim
\frac{\left(  \chi^{0}\right)  ^{2}}{L^{2}}$ becomes comparable with $\phi$
given in (\ref{73}). The contributions of quantum fluctuations to the
gravitational potential from the scalar mode of the graviton $\pi$ and mimetic
matter $\lambda$ are given by (\ref{30a}) and (\ref{30b}) respectively.
Therefore we can easily see that for $\pi$ mode the linearized approximation
for constraint (\ref{72}) remains always valid for these perturbations. For
mimetic mode the last term in (\ref{72}) starts to exceed the linear term at
$L\leq m^{-\frac{3}{5}}$ when $\delta\lambda_{L}$ becomes of order $m^{2}$ and
in case of mimetic matter the linear approximation fails for the constraint
$\bar{h}_{0}^{0}=0$.

Now we calculate higher order corrections to equations (\ref{28})-(\ref{31}).
Equation (\ref{28}) becomes%
\begin{align}
\Delta\psi &  =\lambda+\frac{m^{2}}{4}\left[  \left(  3\psi+\Delta\pi\right)
+\dot{\pi}_{,i}\dot{\pi}_{,i}+\frac{3}{2}\pi_{,ik}\pi_{,ik}-\frac{1}{2}\left(
\Delta\pi\right)  ^{2}-\chi_{,i}^{0}\dot{\pi}_{,i}-\frac{1}{2}\chi_{,i}%
^{0}\chi_{,i}^{0}\right] \nonumber\\
&  +O\left(  \lambda\dot{\chi}^{0},\psi^{2},\psi\Delta\pi,...\right)  ,
\label{77}%
\end{align}
where we have assumed that the gravitational potentials $\psi,$ $\phi$ are
much smaller than unity. We have neglected in (\ref{77}) the higher order
terms in metric dependence in both sides of Einstein equations as well as
higher order terms such as $\psi\Delta\pi$ etc. because they are always
smaller than the corresponding linear terms. Considering perturbations with
wavelength $L\ll m^{-1}$ and taking into account (\ref{73})-(\ref{75}) we can
estimate the various quadratic terms in (\ref{77}) as
\begin{equation}
\dot{\pi}_{,i}\dot{\pi}_{,i}\sim\pi_{,ik}\pi_{,ik}\sim\left(  \Delta
\pi\right)  ^{2}\sim\frac{\pi^{2}}{L^{4}};\text{ \qquad}\chi_{,i}^{0}\dot{\pi
}_{,i}\sim\frac{\lambda\pi}{mL};\text{ \qquad}\chi_{,i}^{0}\chi_{,i}^{0}%
\sim\frac{\lambda^{2}L^{2}}{m^{2}}, \label{78}%
\end{equation}
where we kept only the leading terms. For quantum fluctuations we have%
\begin{equation}
\frac{\pi^{2}}{L^{4}}\sim\frac{m^{2}}{\left(  mL\right)  ^{4}};\qquad\text{
}\frac{\lambda\pi}{mL}\sim\frac{m^{2}}{\left(  mL\right)  ^{7/2}};\text{
\qquad}\frac{\lambda^{2}L^{2}}{m^{2}}\sim\frac{m^{2}}{\left(  mL\right)  ^{3}%
}. \label{79}%
\end{equation}
These nonlinear terms have to be compared to the linear terms in the brackets
on the right hand side of equation (\ref{77})%
\begin{equation}
\Delta\pi\sim\frac{\pi}{L^{2}}\sim\frac{m}{\left(  mL\right)  ^{2}};\text{
\qquad}\psi\sim L^{2}\lambda\sim\frac{m}{\left(  mL\right)  ^{1/2}},
\label{80}%
\end{equation}
where the first term is entirely due to the contribution of $\pi$ mode of the
graviton and the main contribution to the second term comes from mimetic
matter. Because $mL\ll1,$ the term $\frac{\pi^{2}}{L^{4}}$ dominates among
nonlinear corrections for quantum fluctuations and is entirely due to the
scalar mode $\pi$ of the massive graviton. It becomes comparable to $\Delta
\pi$ at the scale%
\begin{equation}
L_{\pi}^{\mathrm{str}}\sim m^{-\frac{1}{2}}, \label{81}%
\end{equation}
and for $L\leq$ $L_{\pi}^{\mathrm{str}}$ the scalar mode of the graviton
becomes strongly coupled and decouples from the two transverse degree of
freedom which, in contrast, only becomes strongly coupled at Planck scale .
Among the two contributions from mimetic matter the second term in (\ref{79})
is obviously larger than the third term and becomes comparable with the linear
term $\psi$ given in (\ref{79}) at the scale%
\begin{equation}
L_{\mathrm{mim}}^{\mathrm{str}}\sim m^{-\frac{2}{3}}, \label{82}%
\end{equation}
where the quantized mimetic matter becomes strongly coupled. This scale is
larger than the scale $\sim m^{-\frac{3}{5}}$ found above, where nonlinear
corrections to the constraint $\bar{h}_{0}^{0}=0$ become important$.$ Hence
the mimetic mode gets in the strongly coupled regime when $\delta\lambda_{L}$
is yet smaller than $m^{2}.$ Notice that at scale $\sim m^{-\frac{3}{5}}$ the
smallest last term in (\ref{79}) becomes of order $\psi$ given in (\ref{80}).
Equation (\ref{29}) gets modified by the dominant nonlinear corrections as%
\begin{equation}
\dot{\psi}_{,i}=-\frac{m^{2}}{4}\left[  \left(  \chi^{0}-\dot{\pi}\right)
_{,i}-\Delta\pi\chi_{,i}^{0}+2\chi_{,k}^{0}\pi_{,ik}+\Delta\pi\dot{\pi}%
_{,i}-3\pi_{,ik}\dot{\pi}_{,k}\right]  +\lambda\chi_{,i}^{0}+... \label{83}%
\end{equation}
For quantum fluctuations the nonlinear terms inside the brackets become
comparable with the linear ones at $L_{\pi}^{\mathrm{str}}\sim m^{-\frac{1}%
{2}}$ both for the scalar mode $\pi$ and quantum mimetic matter. The last term
in this equation $\lambda\chi_{,i}^{0}$ becomes of order $m^{2}\chi_{,i}^{0}$
at scales $L\sim m^{-\frac{3}{5}}$ when the amplitude of quantum fluctuation
of $\delta\lambda$ would reach the value $m^{2}.$ However, as we have seen
above the linearized approximation for mimetic matter fails before, at
$L_{\mathrm{mim}}^{\mathrm{str}}\sim m^{-\frac{2}{3}}.$

The higher order corrections to $i-k$ components of Einstein equations have
the same structure as in (\ref{77}) and hence do not lead to any further
restrictions on the linear perturbation theory. Thus, the scalar mode of the
graviton $\pi$ gets strongly coupled at length scales of order $m^{-\frac
{1}{2}}$ and at smaller scales the massive graviton loses this degree of freedom.

\bigskip

\textbf{Vector modes. }For the vector perturbations equations (\ref{58}) and
(\ref{59}), taking into account the \textit{relevant} quadratic terms, become%
\begin{equation}
\Delta S_{i}=m^{2}\left[  (S_{i}-\dot{\tilde{\chi}}_{i})+2\dot{\tilde{\chi}%
}_{k}\tilde{\chi}_{k,i}+\dot{\tilde{\chi}}_{k}\tilde{\chi}_{i,k}\right]  ,
\label{84}%
\end{equation}%
\begin{align}
\dot{S}_{i,k}+\dot{S}_{k,i}  &  =-m^{2}\left[  \left(  \tilde{\chi}%
_{i,k}+\tilde{\chi}_{k,i}\right)  +\dot{\tilde{\chi}}_{i}\dot{\tilde{\chi}%
}_{k}-\delta_{ik}\dot{\tilde{\chi}}_{m}\dot{\tilde{\chi}}_{m}-\tilde{\chi
}_{i,m}\tilde{\chi}_{k,m}-2\tilde{\chi}_{m,i}\tilde{\chi}_{m,k}\right.
\nonumber\\
&  \left.  -\tilde{\chi}_{i,m}\tilde{\chi}_{m,k}-\tilde{\chi}_{k,m}\tilde
{\chi}_{m,i}+\delta_{ik}\left(  \tilde{\chi}_{l,m}\tilde{\chi}_{l,m}+\frac
{1}{2}\tilde{\chi}_{m,l}\tilde{\chi}_{l,m}\right)  \right]  , \label{85}%
\end{align}
where we have omitted the quadratic terms $S_{i}^{2},$ $S_{i}\dot{\tilde{\chi
}}_{i},$ $S_{m}\tilde{\chi}_{i,m}$ because the metric perturbations $S_{i}$
are always much smaller than unity. For plane-wave perturbations with momentum
$k\simeq L^{-1}\gg m$ these quadratic terms in (\ref{84}), (\ref{85}) can be
estimated as $\frac{\tilde{\chi}^{2}}{L^{2}}$ and these become comparable with
the linear terms $\frac{\tilde{\chi}}{L}$ when $\frac{\tilde{\chi}}{L}\sim1$.
For the minimal quantum fluctuations $\tilde{\chi}\sim\frac{1}{mL}$ (see
(\ref{66})) and hence vector modes of the massive graviton come in the strong
coupling regime at the same scale as the scalar mode $\pi$%
\begin{equation}
L_{\tilde{\chi}}^{\mathrm{str}}\sim m^{-\frac{1}{2}}. \label{86}%
\end{equation}
Thus, for $L\ll m^{-\frac{1}{2}}$ the massive graviton loses three out of five
degrees of freedom and the two remaining transverse degrees of freedom
continue to propagate as if the graviton would be massless. One can easily
check that these two degrees of freedom are not influenced much by scalar and
vector degrees of freedom via nonlinear corrections because the metric
perturbations they induce always remain much smaller than unity. The
transverse degrees of freedom become strongly coupled only at the Planck
scale. When mass of the graviton vanishes, the strong coupling scale for
scalar and vector modes goes to infinity and the graviton has only two
propagating degrees of freedom as it must be.

\bigskip

\section{\bigskip Static gravitational field by external source}

Let us consider the spherically symmetric field created by the mass $M$. In
General Relativity the metric far away from the source (at distance $r$) can
be written in the linear approximation as
\begin{equation}
ds^{2}=(1+2\phi)dt^{2}-\left(  1-2\psi\right)  \delta_{ik}dx^{i}dx^{k},
\label{87}%
\end{equation}
where
\begin{equation}
\phi=\psi=\phi_{N}=-\frac{M}{8\pi r}, \label{88}%
\end{equation}
is the Newtonian gravitational potential (in units $8\pi G=1$).

In massive gravity this potential is modified at distances $r\gg m^{-1}$ where
it decays exponentially fast. However, more nontrivial is that for the case of
Fierz-Pauli mass term Bianchi identities inevitably enforce the scalar
curvature perturbations to vanish in the linear order \cite{BD}:%
\begin{equation}
\delta R=2\Delta\left(  2\psi-\phi\right)  =0, \label{89}%
\end{equation}
from which it follows that $\phi=2\psi.$ The bending of light measures the sum
of two potentials $\phi+\psi,$ that must be equal to $2\phi_{N}.$ Hence, the
gravitational potential $\phi$ determining for instance the motion of the
planets must be in this case equal to $\frac{4}{3}\phi_{N},$ independently of
the mass of the graviton, in contradiction with experiment. This is known as
vDVZ discontinuity which survives in the linearized theory even in the limit
when the mass of the graviton goes to zero \cite{vDVZ}. The resolution to this
apparent contradiction was found by Vainshtein who showed that the scalar mode
of the massive graviton which gives extra $\frac{1}{3}\phi_{N}$ contribution
to the gravitational potential $\phi$ gets strongly coupled at the scale
$R_{V},$ where the linearized approximation for this mode fails and the result
of General Relativity is restored in the leading order for $r<R_{V}$
\cite{Vein}. Depending on the nonlinear extension of Fierz-Pauli mass term,
the Vainshtein scale $R_{V}$ changes in the interval from $\left(  \frac
{M}{m^{4}}\right)  ^{\frac{1}{5}}$ to $\left(  \frac{M}{m^{2}}\right)
^{\frac{1}{3}}$ (see, for example, \cite{ACM}, and references there). When
mass of the graviton goes to zero, Vainshtein scale grows to infinity and thus
the range of scales with vDVZ discontinuity disappears.

In the theory considered here the mass term is not of FP type and vDVZ
discontinuity does not arise even at linearized level. In the presence of an
external source of mass $M,$ among equations (\ref{28})-(\ref{32}), describing
the linearized scalar mode of massive graviton, only equation (\ref{28}) gets
modified as%
\begin{equation}
\Delta\psi=\frac{m^{2}}{4}\left(  3\psi+\Delta\pi\right)  +\lambda+\frac{1}%
{2}T_{0}^{0}, \label{90}%
\end{equation}
where $T_{0}^{0}=M\delta^{3}\left(  x\right)  .$ In the consideration above
this equation was used only to find $\lambda$ for a given $\psi$ and $\pi
.$\ In turn the solution for $\psi$ and $\pi$ is given in (\ref{35}) and
(\ref{37}) and remains unchanged. First of all let us note that $\pi$
completely decouples from mimetic matter and the gravitational field induced
by a central source because $T_{0}^{0}$ comes only in combination with
$\lambda$ and only in equation (\ref{28}). Therefore we can set $\pi=0$
assuming that the constants of integration in (\ref{35}) vanish. Then as
follows from (\ref{37}) and (\ref{38})%
\begin{align}
\phi &  =\psi=C\left(  x^{i}\right)  \sin\frac{mt}{2}+D\left(  x^{i}\right)
\cos\frac{mt}{2}\nonumber\\
&  =F\left(  x^{i}\right)  \cos\left(  \frac{mt}{2}+\alpha\left(
x^{i}\right)  \right)  . \label{91}%
\end{align}
It is interesting to note that in case of massive mimetic gravity the static
solutions for the static source do not exist in the linearized version of the
theory. The gravitational potential oscillates with a frequency proportional
to the mass of the graviton. Of course on time scales $t\ll m^{-1}$ the
potential does not change too much. Considering $t\ll m^{-1}$ and scales $r\ll
m^{-1}$ we can estimate the constant of integration in (\ref{91}) in the
presence of source $T_{0}^{0}$ from (\ref{90}) to get%
\begin{equation}
\phi=\psi\simeq-\frac{1}{8\pi\cos\alpha}\left(  \frac{M}{r}+\frac{2}{r}%
\int^{r}\lambda d^{3}x\right)  \cos\left(  \frac{mt}{2}+\alpha\right)  ,
\label{92}%
\end{equation}
and as follows from (\ref{29})%
\begin{equation}
\chi^{0}=-\frac{4}{m^{2}}\dot{\psi}\simeq-\frac{1}{4\pi m\cos\alpha}\left(
\frac{M}{r}+\frac{2}{r}\int^{r}\lambda d^{3}x\right)  \sin\left(  \frac{mt}%
{2}+\alpha\right)  \label{93}%
\end{equation}
Clearly these solutions are approximate solutions valid only for $t\ll
m^{-1}.$ Keeping this in mind we will find the range of the scales where the
linear approximation used to get them fails because of nonlinear corrections.
From (\ref{72}) and (\ref{77}) it is clear that this happens when $\chi
_{,i}^{0}\chi_{,i}^{0}$ becomes of order $\phi=\psi.$ Neglecting the
contribution of mimetic matter in (\ref{92}), (\ref{93}) and taking as a rough
estimate%
\begin{equation}
\psi\sim\frac{M}{r},\text{ \ \ \ }\chi_{,i}^{0}\chi_{,i}^{0}\sim\frac{\left(
\chi^{0}\right)  ^{2}}{r^{2}}\sim\frac{M^{2}}{m^{2}r^{4}}, \label{94}%
\end{equation}
we see that these terms become comparable at%
\begin{equation}
r_{\mathrm{str}}\sim\left(  \frac{M}{m^{2}}\right)  ^{\frac{1}{3}} \label{96}%
\end{equation}
and for $r<r_{\mathrm{str}}$ the nonlinear term $\chi_{,i}^{0}\chi_{,i}^{0}$
cannot be neglected. Thus the linearized approximation used to derive the
non-static solutions (\ref{92}), (\ref{93}) is not valid for
$r<r_{\mathrm{str}}$. Moreover as follows from (\ref{83})%
\begin{equation}
\dot{\psi}_{,i}=\left(  \lambda-\frac{m^{2}}{4}\right)  \chi_{,i}^{0},
\label{97}%
\end{equation}
and hence when $\lambda$ becomes comparable with $m^{2}$ the linearized $0-i$
equations fail. One can see that this equation allows static solution for
$\lambda=m^{2}/4.$ In this case the contribution of mimetic matter to the
gravitational potential in (\ref{92}) becomes comparable to the contribution
of the source of mass $M$ at scales $r_{\mathrm{str}}\sim\left(  \frac
{M}{m^{2}}\right)  ^{\frac{1}{3}}.$

Thus, at scales $r<r_{\mathrm{str}}$ we find that the solution of the equation
for central source is static and corresponds to%
\begin{equation}
\phi=\psi=\phi_{N}\left(  1+O\left(  1\right)  \left(  \frac{r}%
{r_{\mathrm{str}}}\right)  ^{3}\right)  , \label{98}%
\end{equation}
where the corrections due to the contribution of induced mimetic matter
rapidly decrease towards the smaller scales. On large scales, for
$m^{-1}>r>r_{\mathrm{str}}$ the static source produces the time dependent
oscillating solution%
\begin{equation}
\phi=\psi\simeq O\left(  1\right)  \phi_{N}\cos\left(  \frac{mt}{2}%
+\alpha\right)  . \label{99}%
\end{equation}
These time dependent oscillations are due to the induced mimetic matter which
surrounds the static source.

\section{Summary}

It is a rather interesting theoretical challenge to try to construct a theory
of massive gravity free of obvious flaws in which, on Minkowski background,
the graviton behaves like a massive particle. This is the main motivation for
this work. Generically the five degrees of freedom of a massive graviton are
accompanied by an extra ghost degree of freedom which propagates already at
the linear level unless the mass term is taken to be in the Fierz-Pauli form
\cite{FP}. For the FP term the dangerous mode remains non-dynamical in the
linear approximation but generically reappears as nonlinear Boulware-Deser
ghost \cite{BD} on non-trivial (different from Minkowski) backgrounds. There
is a claim in the literature that there exists a nearly unambiguous higher
order extension of the FP theory in which such ghost is absent on most of
\ the non-trivial backgrounds \cite{Kab}. However, as we have shown in
\cite{CMghost,CMghost1}, the ghost field can still reappear on some
backgrounds . In this paper we have addressed the following questions:

\begin{itemize}
\item \textit{Is it possible to have a rather simple theory of massive
gravity, which remains ghost free irrespective of its nonlinear extension}?

\item \textit{Can we have non FP mass term and nevertheless avoid the
appearance of a ghost? }
\end{itemize}

As we have shown in this paper the answer to both questions is positive. The
only way to generate the graviton mass without explicitly spoiling
diffeomorphism invariance is to use the BEH mechanism with four scalar fields.
In this case, in the broken symmetry phase, the three scalar fields
\emph{donate}\ their degrees of freedom to the graviton which thus gets a
total of five degrees of freedom and acquires mass. The fourth field, if
unrestricted, becomes a ghost and even for the Fierz-Pauli mass form still
propagates on some non-trivial backgrounds. We have proposed to use the
constrained mimetic scalar instead of this dangerous field. This field becomes
dynamical and can imitate dark matter when combined with the longitudinal mode
of gravity. In mimetic massive gravity its behavior is slightly modified and
mimetic matter still well imitates cold dark matter. Thus, the dangerous mode
is avoided and instead we get a good candidate for dark matter irrespective of
the nonlinear extension of the theory. The inevitable consequence of using
mimetic field, which is always in \emph{broken symmetry phase},\ is the need
to use non Fierz-Pauli type mass term, with a relative coefficient of
$-\frac{1}{2}$ between the $\bar{h}^{2}$ and $\bar{h}_{B}^{A}\bar{h}_{A}^{B}$
terms instead of $-1.$ Only in this case we obtain a massive graviton in the
broken symmetry phase. This graviton is completely decoupled from mimetic
matter in the linear approximation and vDVZ discontinuity is completely
avoided. Out of the five degrees of freedom, three degrees due to the scalar
fields get in the strongly coupled regime at the same scale $L^{str}\sim
m^{-\frac{1}{2}}$ and at $L<L^{\mathrm{str}}$ the graviton has only two
transverse degrees of freedom which become strongly coupled at Planck scale.
This is quite different from the case of FP mass term where these scales are
different for the scalar and vector modes. The other interesting feature of
the theory is that the gravitational potential produced by a mass $M$ is
static only at scales $L<$ $r_{\mathrm{str}}\sim\left(  \frac{M}{m^{2}%
}\right)  ^{\frac{1}{3}}$ while at larger distances it oscillates with
frequency $\frac{m}{2}$ and thus vanishes, being averaged over the time scales
$t>m^{-1}.$ \vspace{2cm}

\textbf{{\large {Acknowledgments}}} \vspace{0.5cm}

The work of A. H. C is supported in part by the National Science Foundation
Grant No. Phys-1518371. The work of V.M. is supported in part by \textit{The
Dark Universe}\ and the Cluster of Excellence EXC 153 \textit{Origin and
Structure of the Universe}. V.M. thanks ENS, where a part of this work is
completed, for hospitality.

\end{document}